\documentclass[twocol]{ametsoc}
\pdfoutput=1 

\usepackage{verbatim} 

\usepackage{flushend} 

\journal{jpo}

\bibpunct{(}{)}{;}{a}{}{,}

\title{Are rogue waves really unexpected?}

\authors{Francesco Fedele\correspondingauthor{Georgia Institute of Technology
Atlanta, GA 30332, USA.}}

\affiliation{School of Civil and Environmental Engineering, School of Electrical and Computer Engineering, Georgia Institute of Technology, Atlanta, GA, USA.}

\email{fedele@gatech.edu}

\abstract{An unexpected wave is defined by~\cite{Gemmrich2008} as a wave that is much taller than a set of neighboring waves. Their definition of "unexpected" refers to a wave that is not anticipated by a casual observer. Clearly, unexpected waves defined in this way are predictable in a statistical sense. They can occur relatively often with a small or moderate crest height, but large unexpected waves that are rogue are rare. Here, this concept is elaborated and statistically described based on a third-order nonlinear model. In particular, the conditional return period of an unexpected wave whose crest exceeds a given threshold is developed. This definition leads to greater return periods or on average less frequent occurrences of unexpected waves than those implied by the conventional return periods not conditioned on a reference threshold. Ultimately, it appears that a rogue wave that is also unexpected would have a lower occurrence frequency than that of a usual rogue wave. As specific applications, the Andrea and WACSIS rogue wave events are examined in detail. Both waves appeared without warning and their crests were nearly $2$-times larger than the surrounding $O(10)$ wave crests, and thus unexpected. The two crest heights are nearly the same as the threshold~$h_{0.3\cdot10^{6}}\sim1.6H_{s}$ exceeded on average once every~$0.3\cdot 10^{6}$ waves, where $H_s$ is the significant wave height. In contrast, the Andrea and WACSIS events, as both rogue and unexpected, would occur slightly less often and on average once every~$3\cdot10^{6}$ and~$0.6\cdot10^6$ waves respectively.}  

\begin{document}

\maketitle

%








\section{Introduction}

A rogue wave is defined as such if the crest-to-trough height is at least $2.2$ times the significant
wave height $H_{s}$ or if the crest height exceeds the threshold $1.25H_{s}$, where
$H_{s}=4\sigma$ and $\sigma$ is the standard deviation of surface
elevations~\citep{DystheKrogstad2008}. Evidences given for the occurrence of such waves in nature include the Draupner and Andrea events.
In particular, the Andrea wave was measured on November 9 2007 by a LASAR system mounted on the Ekofisk platform in the North Sea in a water depth of $d=74$ m \citep{Magnusson2013}.
The Draupner freak wave was measured by Statoil at a nearby platform in January 1995 \citep{haver2001evidences}. In the last decade, the properties of the Draupner and Andrea waves
have been extensively studied~(\cite{DystheKrogstad2008,Osborne1995,Magnusson2013,Bitner_Andrea2014,Dias2015} and references therein). 

\begin{figure*}[th]
\centering\includegraphics[width=1\textwidth]{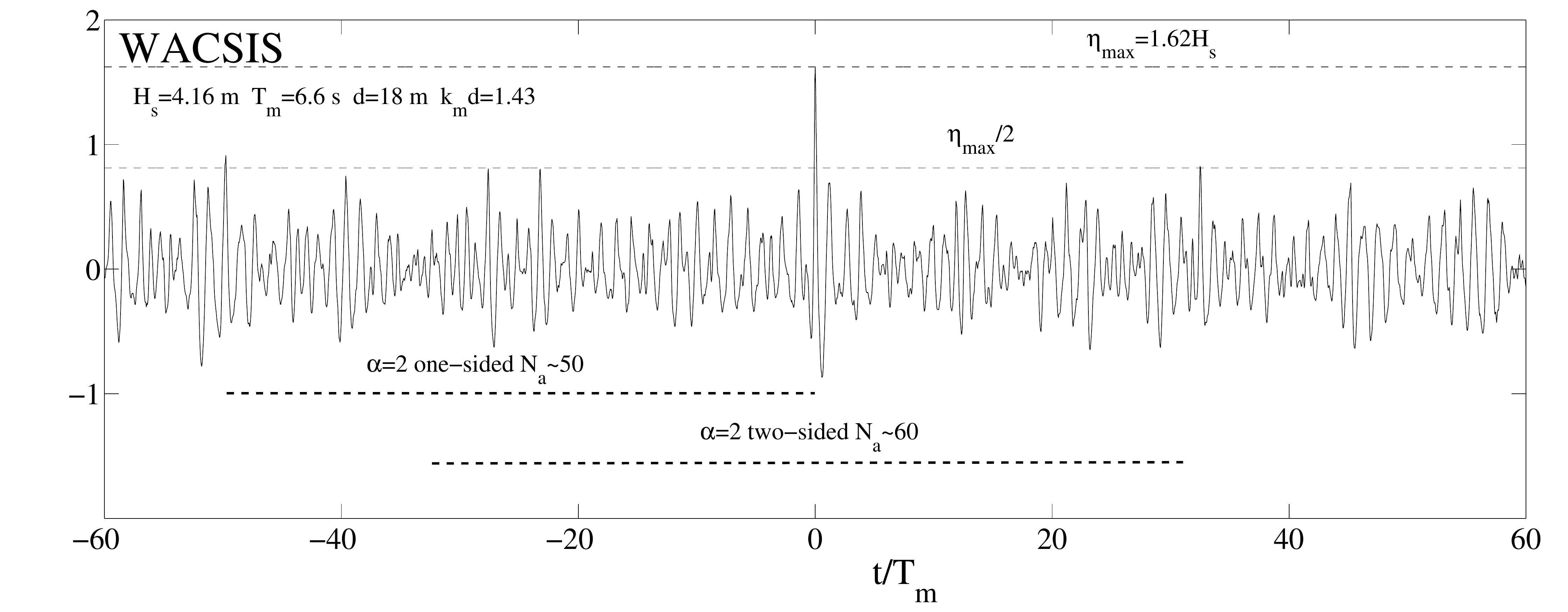} \protect\caption{WACSIS measurements: the observed largest crest height is $\alpha=2$-times
larger than the crests of the one-sided (two-sided) $N_{a}\sim50$
($60$) waves. Wave parameters $H_{s}=4.16$~m, $T_{m}=6.6$~s, depth~$d=18$ m~\citep{WACSIS2002}.}\label{FIG1} 
\end{figure*}

The Andrea wave occurred during a sea state with significant wave height $H_{s}=4\sigma=9.2$
m, mean period $T_{0}=13.2$~s and wavelength $L_{0}=220$~m. The Andrea crest
height is $h=1.63H_s=15$~m and the crest-to-trough height
$H=2.3H_s=21.1$~m.  The sea state during which the Draupner wave occurred had a significant wave height $H_{s}=11.9$ m, mean period $T_{0}=13.1$ s and wavelength
$L_{0}=250$ m. The Draupner crest height is $h=18.5$ m ($h/H_{s}=1.55$)
and the associated crest-to-trough height $H=25.6$ m ($H/H_{s}=2.15$) (\cite{haver2004possible,Magnusson2013}). 
Observations of such large extreme
waves show that they tend to extend above the surrounding smaller waves without warning and thus unexpectedly. Further, both waves were twice as high as the immediately preceding as well as following groups of waves.
In describing the unexpectedness of ocean waves, \cite{Gemmrich2008} define as unexpected a wave $\alpha$-times larger than a set of one-sided (preceding) waves or two-sided (preceding and following) waves (see Fig.~\ref{FIG1}). Note that their definition of unexpectedness refers to the time interval of apparent calm before or during which a wave is much taller than the neighboring waves. Hereafter, the term "unexpected" refers to a wave that is not anticipated by a casual observer as emphasized by~\cite{Gemmrich2010}.  Clearly, unexpected waves defined in this way are predictable in a statistical sense as one can estimate the associated return period or frequency of occurrence.

Indeed, unexpected waves occur often with a small or average wave height, but they are rarely the largest waves in a record or rogue waves~\citep{Gemmrich2010}. 
In this regard,~\cite{Gemmrich2008} performed Monte Carlo simulations of second order nonlinear seas characterized with the typical JONSWAP ocean spectrum and initial homogeneous random conditions. They estimated that a wave with height at least twice that of any of the preceding $30$ waves occurs once every $10^{5}$ waves on average. 
Also unexpected crest heights are more probable than unexpected wave heights as they occur on average once every $7\cdot10^{4}$ in Gaussian seas and once every $10^{4}$ waves in second-order nonlinear seas (see Fig.~2 in~\cite{Gemmrich2008}). Thus, their numerical predictions indicate that in weakly nonlinear seas unexpected waves occur frequently and more often than in Gaussian seas. 

Further,~\cite{Gemmrich2008} noted in their simulations that among the unexpected waves $2$-times larger than the surrounding $30$ waves, only about $q=10-20\%$ were rogues. With reference to second order crest heights, this means that in a sample population of $10^6$ waves a set of $100$ waves are unexpected, as they occur once every $10^4$ waves on average. However, only about $10-20$ waves of the set have crest heights that are rogue, i.e. larger than $1.34H_s$ as the rogue threshold adopted by~\cite{Gemmrich2008}. This implies that unexpected wave crests that are rogue would occur less often, i.e. once every~$10^5$ waves on average. Further, the percentage $q$ of rogue occurrences can be interpreted as the probability that the crest of an unexpected wave exceeds the threshold $1.34H_s$. Consequently, unexpected crest heights larger than $1.34H_s$ would occur rarely.

The preceding results 
provide the principal motivation here to consider a statistical model for describing unexpected waves and their rogueness. We will show that~\cite{Gemmrich2008}'s definition of return period is unconditional. In particular, it is the harmonic mean of the return periods of all unexpected waves with any amplitude. Thus, unexpected waves of moderate amplitude occur relatively often. However, unexpected waves that are rogue have a lower occurrence frequency, and this is in agreement with~\cite{Gemmrich2008}'s numerical predictions.

The remainder of the paper is structured as follows. First, we introduce a new theoretical model for the statistics of unexpected waves that accounts for both second- and third-order nonlinearities. We also study the effects of nonstationarity and stochastic dependence among successive waves.
In particular, we present analytical solutions for the return period of unexpected
waves and associated unconditional and conditional averages for crest
and wave heights. Then, the conceptual framework is validated
by way of Monte Carlo simulations and the theoretical
predictions are compared to oceanic measurements. As a specific
application here, we capitalize on the numerical simulations of the
Andrea sea state \citep{Bitner_Andrea2014,Dias2015} and examine the unexpectedness of the Andrea wave in detail. Summary and conclusions follow subsequently.

\begin{figure*}[th]
\centering\includegraphics[width=1\textwidth]{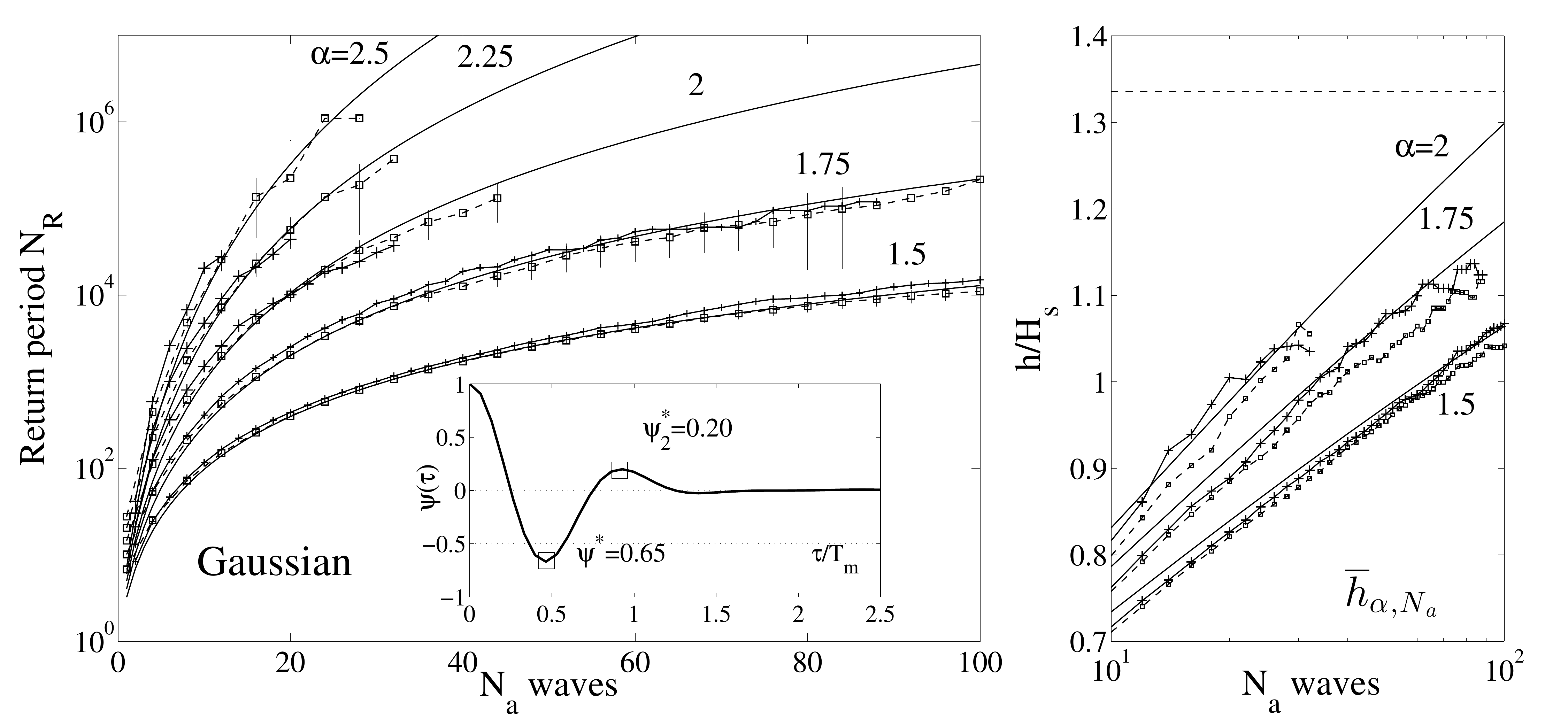}
\protect\caption{Unexpected crest heights in broadbanded Gaussian seas. Left panel: empirical one-sided (thin dashed line with~$\square$) and two-sided (thin solid line with~$+$), $N_{a}$ even) unexpected wave statistics versus (solid line) predicted theoretical unconditional return period $N_{R}$ in number of waves of a wave whose crest height is $\alpha$-times larger than the surrounding $N_{a}$ waves for increasing values of $\alpha=1.5,2$ and $2.5$. Confidence bands are also shown. Right panel: empirical one-sided (thin dashed line with~$\square$) and two-sided (thin solid line with~$+$, $N_{a}$ even) unexpected wave statistics versus
theoretical predictions (solid line) of the mean crest height $\overline{h}_{\alpha,N_{a}}$ of a wave whose crest height is~$\alpha$-times larger than surrounding
$N_{a}$ waves for $\alpha=1.5,1.75,$ and $2$. Sea state parameters: fully developed JONSWAP spectrum (peak enhancement factor $\gamma=1$), mean period $T_{m}=8.3$ s, spectral bandwidth $\nu=0.35$, Boccotti parameters $\psi^{*}=0.65$, $\psi^{*}_2=0.20$ and simulated $\sim10^{6}$ waves (see left panel inset). The theoretical predictions accounting for the stochastic independence and dependence of successive crest heights are practically the same as the sea state is broadbanded.}\label{FIG2} 
\end{figure*}

\section{Statistics of unexpected waves}

Consider the exceedance probability distribution of wave crests characterized by third-order nonlinearities and described by~\citep{TayfunFedele2007}
\begin{equation}
P(x)=\mathrm{Pr}\left[h>x\,H_s\right]=\mathrm{exp}\left(-8\,x_{0}^{2}\right)\left[1+\varLambda x_{0}^{2}\left(4\,x_{0}^{2}-1\right)\right],\label{Pid}
\end{equation}
where $x=h/H_s$ is the crest amplitude $h$ scaled by the significant wave height $H_s=4\sigma$ and $x_{0}$ follows from the quadratic equation \citep{Tayfun1980} 
\begin{equation}
x=x_{0}+2\mu\,x_{0}^{2}.
\end{equation}
Here, the wave steepness $\mu=\lambda_{3}/3$ relates to the
skewness of surface elevations \citep{Fedele2009} and the parameter 
\begin{equation}
\varLambda=\lambda_{40}+2\lambda_{22}+\lambda_{04}\label{gam}
\end{equation}
is a measure of third-order nonlinearities as a function of the fourth
order cumulants $\lambda_{nm}$ of the wave surface $\eta$ and its
Hilbert transform $\hat{\eta}$ \citep{TayfunFedele2007}. \cite{Janssen2006} assume the following relations between cumulants 
\begin{equation}
\lambda_{22}=\lambda_{40}/3,\qquad\lambda_{04}=\lambda_{40},\label{cum}
\end{equation}
which, to date, have been proven to hold for second-order narrowband waves only \citep{TayfunLo1990}. Then, $\varLambda$ in Eq.~\eqref{gam} is approximated in terms of the excess kurtosis $\lambda_{40}$ by
\begin{equation}
\varLambda_{\mathrm{appr}}=\frac{8\lambda_{40}}{3},\label{gamma}
\end{equation}
which will be used in this work. Then, Eq.~\eqref{Pid} reduces to
a modified Edgeworth-Rayleigh (MER) distribution \citep{Janssen2006}.
For realistic oceanic seas the kurtosis $\lambda_{40}$ is mainly affected by bound nonlinearities \citep{Shrira2014_JPO,fedele2014kur,fedele2015JPO}.

Consider now a time interval $\mathcal{T}$ during which
a stationary sequence of $N_{w}=\mathcal{T}/T_{m}$ consecutive waves occur
on average. We assume that neighboring waves are stochastically independent.
This assumption is convenient for the theoric development of a probabilistic model.
Furthermore,~\cite{Borgman} argues that~"\textit{... It seems reasonable to
assume that a wave height is at most interdependent with the first
several wave heights occurring before and after it and essentially
independent with waves further back into the past or forward into
the future}''. We will show later that this is justified as long as
the sea state is broadbanded so that the covariance function decays sufficiently rapid 
to zero after few wave periods and successive wave peaks decorrelate faster.
Thus, in a sample of $N_{a}+1$ successive waves it is irrelevant what wave is the unexpected wave larger than the surrounding waves. Indeed, any wave in the sample could be "$p$-sided" unexpected, i.e. $\alpha$-times larger than the previous $m$ waves and following $N_{a}-m$ waves, with $m=1,...N_{a}/2$
and $p=N_{a}/m$. For instance, the last wave in the sample could
be larger than the preceding (one-sided) $N_{a}$ waves ($m=N_{a}$ and $p=1$), or the central wave could extend above the preceding and following (two-sided) $m=N_{a}/2$ waves ($p=2$ and $N_{a}$ even) (see Fig.~\ref{FIG1}). Note that our definition of two-sided unexpectedness is different than that in~\cite{Gemmrich2008} as they consider $N_a$ waves on each side.

Clearly, the statistics of one- and two-sided unexpected
waves, or more generally the $p$-sided statistics are the same if
stochastic independence of successive waves holds. On this basis,
the fraction of waves $n(x;\alpha,N_{a})$ that have a dimensionless
crest height $h/H_s$ within the interval $(x,x+dx)$ and that is $\alpha$-times larger than any of the surrounding $N_{a}$ waves is given by
\begin{equation}
n(x;\alpha,N_{a})dx=\left[1-P\left(\frac{x}{\alpha}\right)\right]^{N_{a}}p(x)dx,\label{nx}
\end{equation}
where $P(x)$ is the exceedance probability given in Eq.~\eqref{Pid} and
\begin{equation}
p\left(x\right)=-\frac{dP}{dx}\label{pdf}
\end{equation}
is the pdf of $x$. Then the probability that the crest height $\xi$ is in $(x,x+dx)$ follows as
\begin{equation}
p_{h}(x;\alpha,N_{a})dx=\frac{n(x;\alpha,N_{a})dx}{n(\alpha,N_{a})},\label{pdfh}
\end{equation}
where $n(\alpha,N_{a})$ is the fraction of waves whose crest height is $\alpha$-times larger than the surrounding $N_{a}$ waves, namely 
\begin{equation}
n(\alpha,N_{a})=\int_{0}^{\infty}n(x;\alpha,N_{a})dx=\int_{0}^{\infty}\left[1-P\left(\frac{x}{\alpha}\right)\right]^{N_{a}}p(x)dx\mbox{.}\label{na}
\end{equation}
By definition, the unconditional return period $R$ or the average time interval between two consecutive occurrences of the unexpected wave event $\mathcal{E}$ is
\begin{equation}
R(\alpha,N_{a})=\frac{\tau}{N_{w}n(\alpha,N_{a})}=\frac{N_{w}T_{m}}{N_{w}n(\alpha,N_{a})}=\frac{T_{m}}{n(\alpha,N_{a})}.\label{R}
\end{equation}
Since $T_{m}$ is the mean wave period, $\mathcal{E}$ occurs on average once every $N_{R}$ waves where 
\begin{equation}
N_{R}(\alpha,N_{a})=\frac{1}{n(\alpha,N_{a})}.\label{NR}
\end{equation}

Another statistical interpretation of the unconditional return period $N_R$ is as follows. Consider the average number of unexpected waves $n_{j}(\alpha,N_{a})\Delta x$
with a crest height between $x_{j}-\Delta x/2$ and $x_{j}+\Delta x/2$,
where $\Delta x\ll1$ is small and $x_{j}$ are increasing amplitudes starting from $x_{1}=0$,
i.e. $x_{j+1}>x_{j}$, for $j=1,\dots$. Then, 
\[
N_{R,j}(\alpha,N_{a})=\frac{1}{n_{j}(\alpha,N_{a})\Delta x}
\]
is the return period of an unexpected wave whose crest height is nearly~$x_{j}$.
Then, Eq.~\eqref{NR} is approximated as
\begin{equation}
N_{R}(\alpha,N_{a})\simeq\frac{1}{\sum_{j=1}^{\infty}n_{j}(\alpha,N_{a})\Delta x}=\frac{1}{\sum_{j=1}^{\infty}\frac{1}{N_{R,j}(\alpha,N_{a})}},
\label{harmonicmean}
\end{equation}
which reveals that $N_{R}$ is the harmonic mean of the return periods $N_{R,j}$ of all unexpected waves with any crest height. 

The associated mean crest height of a wave $\alpha$-times larger than the surrounding $N_{a}$ waves follows from Eq.~\eqref{pdfh} as 
\begin{equation}
\overline{h}_{\alpha,N_{a}}=H_s\int_{0}^{\infty}xp_{h}(x;\alpha,N_{a})dx.\label{hmean}
\end{equation}

For comparison purposes, we also consider the standard statistics $\overline{h}_{\mathrm{max},n}$, $h_{n}$ and $h_{1/n}$ for crest heights (\cite{TayfunFedele2007}). In particular, $\overline{h}_{\mathrm{max},n}$ is the mean maximum crest height of a sample of $n$ waves
\begin{equation}
\overline{h}_{\mathrm{max},n}=H_s\int_{0}^{\infty}\mathit{\mbox{\ensuremath{\left\{ 1-\left[1-P\left(x\right)\right]^{n}\right\} dx},}}\label{hmaxn}
\end{equation}
which admits Gumbel-type asymptotic approximations (\cite{TayfunFedele2007,fedele2015JPO}).
Further, $h_{n}$ is the threshold exceeded by the $1/n$ fraction
of largest crest heights and it satisfies
\begin{equation}
P(h_n/H_s)=\frac{1}{n},\label{Phn}
\end{equation}
where $P(x)$ is the unconditional nonlinear probability of exceedance for
crest heights given in Eq.~\eqref{Pid}. The statistics~$h_{1/n}$ is the
conditional mean $\overline{h\left|h>h_{1/n}\right.}$, namely the
average of the $1/n$ fraction of largest crest heights
\begin{equation}
h_{1/n}=h_{n}+n H_s\int_{h_{n}}^{\infty}P(x)dx.\label{h1n}
\end{equation}
One can show that $\overline{h}_{\mathrm{max},n}$ is always smaller than $h_{1/n}$ and they tend to be the same as $n$ increases~\citep{TayfunFedele2007}.

We also consider the standard conditional return period $N_h(\xi)$ (in number of waves) of a wave whose crest exceeds the threshold $h=\xi H_s$, namely
\begin{equation}
N_h(\xi)=\frac{1}{\mathrm{Pr}\left[h>\xi H_s\right]}=\frac{1}{P(\xi)},\label{Nh}
\end{equation}
where the exceedance probability $P(\xi)$ is that in~Eq.\eqref{Pid}.  From Eq.~\eqref{Phn}, the threshold~$h_n$ exceeded with probability $1/n$ implies that $N_h(h_n/H_s)=n$, i.e. on average $h_n$ is  exceeded once every $n$ waves. 

Similar statistics for the crest-to-trough height $y=H/H_s$ of unexpected waves follow by replacing the crest exceedance probability $P$ in Eq.~\eqref{Pid} with the generalized Boccotti distribution \citep{Tayfun2013_GenBoccotti} 
\begin{equation}
\begin{split}
&P_{H}(y)=\mathrm{Pr}\left[H>y\,H_s\right]=\\
&c_{0}\mathrm{exp}\left(-\frac{4\,y^{2}}{1+\psi^{*}}\right)\left[1+\frac{\varLambda\,y^{2}}{1+\psi^{*}}\left(\frac{y^{2}}{1+\psi^{*}}-\frac{1}{2}\right)\right],\label{PH}
\end{split}
\end{equation}

where 
\[
c_{0}=\frac{1+\ddot{\psi}^{*}}{\sqrt{2\,\ddot{\psi}^{*}\left(1+\psi^{*}\right)}},
\]
and $\psi^{*}=\psi(\tau^{*})$ is the absolute value of the first minimum of the normalized covariance function $\psi(\tau)=\overline{\eta(t)\eta(t+\tau)}/\sigma^{2}$ of the zero-mean random wave process $\eta(t)$, which is attained at $\tau=\tau^{*}$ and $\ddot{\psi}^{*}$ the corresponding second derivative \citep{Boccotti2000}. 

The corresponding linear statistics of unexpected wave crests follow by setting $\mu=0$ and $\Lambda=0$ in Eq.~\eqref{Pid}, or $\Lambda=0$ in Eq.~\eqref{PH} for wave heights. These will hereafter be differentiated with the superscript $L$. In the following, we will not dwell that much on unexpected wave heights, but our main focus will be the statistics of unexpected crests in typical
oceanic sea states. 

Finally, we point out that our present theory of unexpected waves can be generalized to space-time extremes drawing on~\cite{Fedele2012}, but this is beyond the scope of this paper. 
\begin{figure}[th]
\centering\includegraphics[width=1\columnwidth]{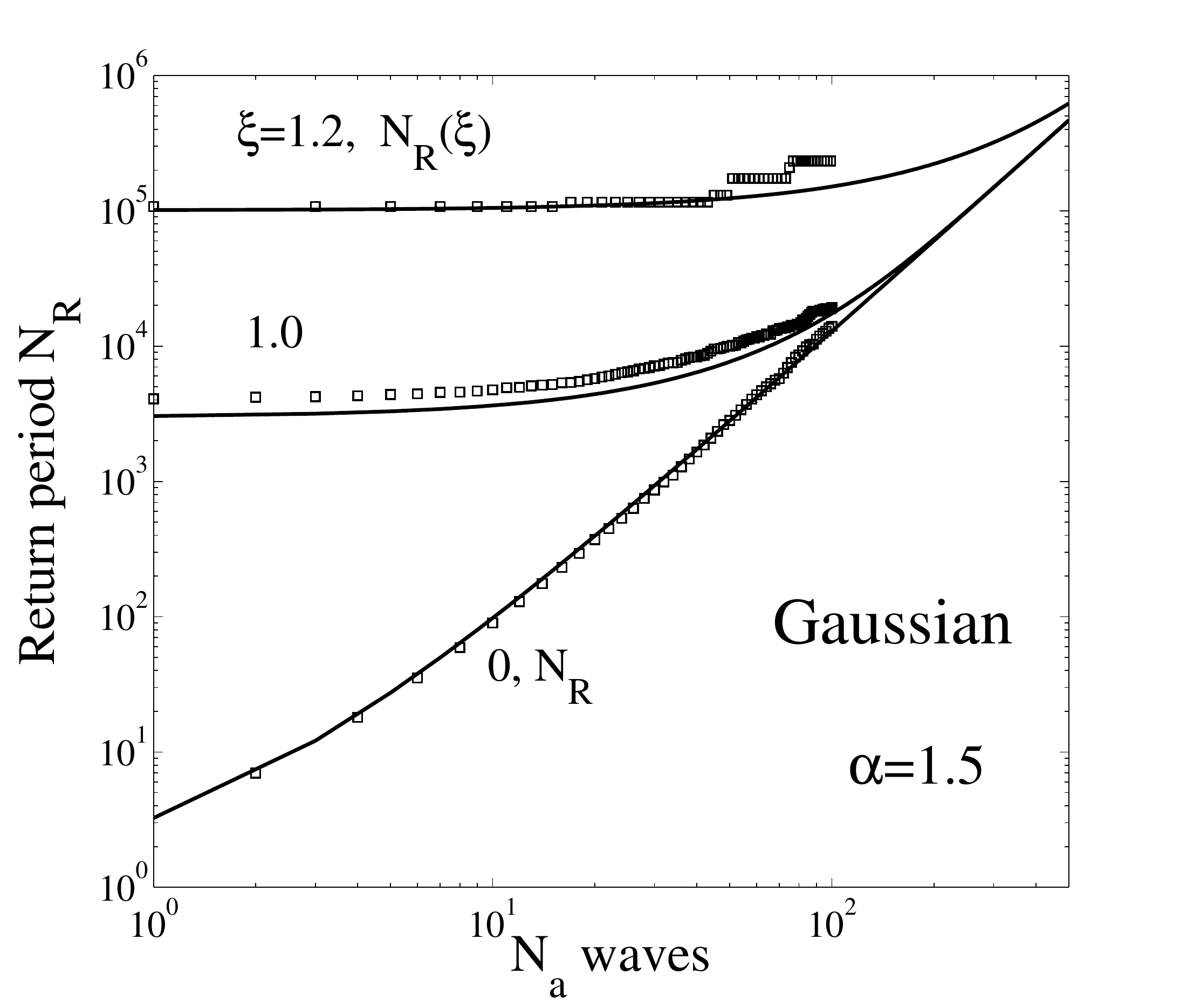}
\protect\caption{Conditional return period of large unexpected waves in Gaussian seas: (square) empirical one-sided unexpected wave statistics versus (solid lines)
predicted theoretical conditional return periods $N_R(\xi)$ in number of waves of unexpected waves whose crest height is greater than $\xi H_s$ and $\alpha=1.5$-times larger than the surrounding $N_{a}$ waves for  $\xi=0,1.0$ and $1.2$. Note that~$N_R(\xi=0)$ is the unconditional return period~$N_R$. Sea state parameters: fully developed JONSWAP spectrum (peak enhancement factor $\gamma=1$), mean period $T_{m}=8.3$ s, spectral bandwidth $\nu=0.35$, Boccotti parameters $\psi^{*}=0.65$, $\psi^{*}_2=0.20$ and simulated $\sim10^{6}$ waves. The predictions accounting for the stochastic independence and dependence of successive crest heights are practically the same as the sea state is broadbanded.}\label{FIG3} 
\end{figure}

\begin{figure*}[th]
\centering\includegraphics[width=1\textwidth]{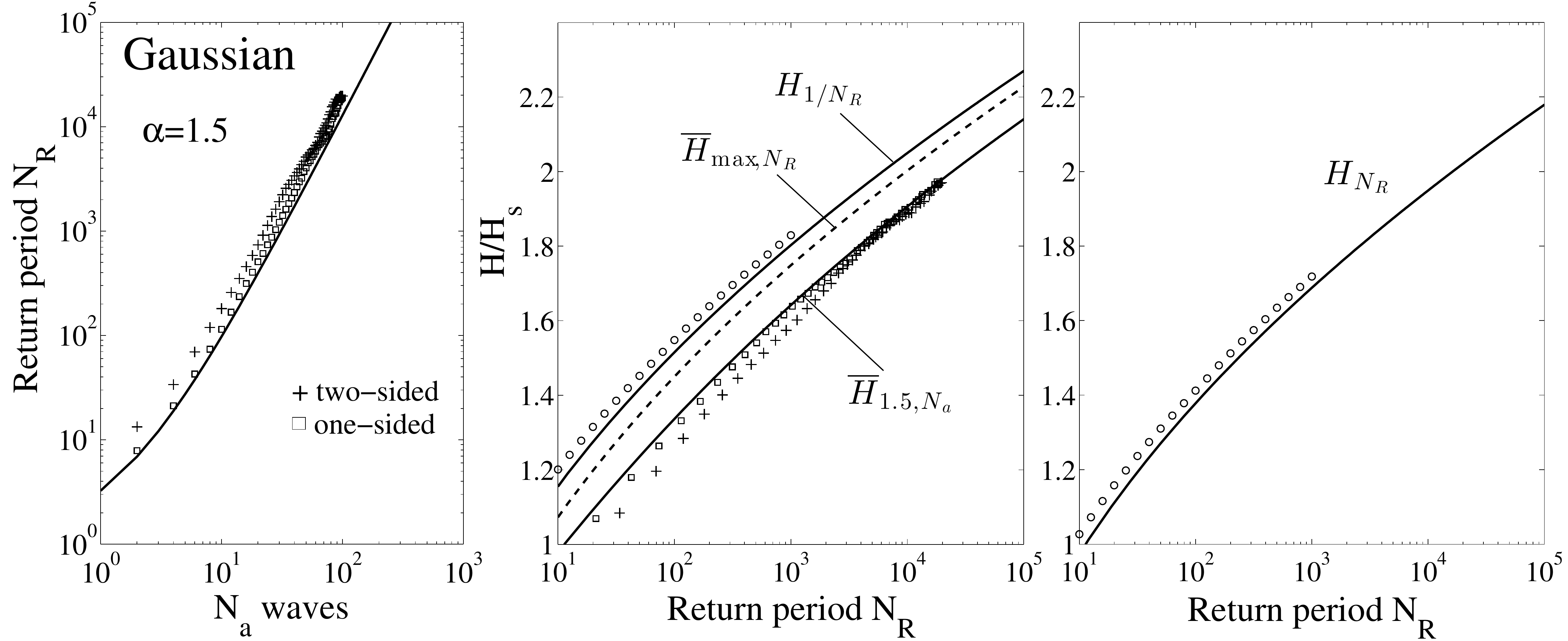}
\protect\caption{Unexpected wave heights in Gaussian seas: (Left panel) predicted theoretical unconditional return period $N_{R}$ in number of waves (solid line) versus empirical one-sided ($+$) and two-sided ($\square$,~$N_{a}$ even) statistics as a function of the number $N_{a}$ of
surrounding waves for $\alpha=1.5$; (center panel) predicted
mean unexpected wave height $\overline{H}_{1.5,N_{a}}$ versus observations as a function
of the return period $N_{R}$.  For comparison purposes, predicted mean wave height $\overline{H}_{\mathrm{max},N_{R}}$, conditional mean $H_{1/N_{R}}$ and (right panel) threshold $H_{N_{R}}$ versus observations (circles) are also shown. 
Sea state parameters: fully developed JONSWAP spectrum (peak enhancement factor $\gamma=1$), mean period $T_{m}=8.3$ s, spectral bandwidth $\nu=0.35$, Boccotti parameter $\psi^{*}=0.65$ and simulated $\sim10^{6}$ waves.}\label{FIG4} 
\end{figure*}

\subsection{Stochastic dependence of successive waves}

The statistics of unexpected waves presented so far does not take
into account the stochastic dependence of neighboring waves or wave groupness. Clearly,
for large crest heights, as argued by~\cite{Borgman}, one expects
that only few neighboring crests are more or less correlated~\citep{watson1954}. To quantify
this, we draw on~\cite{Fedele2005} and model a stationary sequence of wave crests $\left\{ x_{j}=\frac{h_{j}}{Hs}\right\} _{j=1,N_{w}}$
as a one-step memory Markov chain, where each crest height $x_j$ is only stochastically
dependent on the preceding crest height $x_{j-1}$, that is 
\[
p(x_{j}|x_{j-1},x_{j-2},...x_{2},x_{1})=p(x_{j}|x_{j-1}).
\]
Since the sequence is stationary, the conditional pdf $p(x_{j}|x_{j-1})$
is the same for any $j$, say $p(x_{2}|x_{1})=p(x_{1},x_{2})/p(x_{1})$,
where the crest $x_{1}$ precedes $x_{2}$, and $p(x_{1},x_{2})$
and $p(x_{1})$ are the associated joint and marginal pdfs.

On these assumptions, following~\cite{Fedele2005} the fraction of waves $n(x;\alpha,N_{a})$
that have a dimensionless crest height within the interval $(x,x+dx)$
and that is $\alpha$-times larger than any of the surrounding $N_{a}$
waves is given by
\begin{equation}
\begin{split}
n(x;\alpha,N_{a})dx=\left[\Pr\left(\left.x_{2}<\frac{x}{\alpha}\right|x_{1}<\frac{x}{\alpha}\right)\right]^{N_{a}-1}\\
\cdot\Pr\left(\left.x_{2}<\frac{x}{\alpha}\right|x_{1}=x\right)\mbox{\mbox{\mbox{\ensuremath{\mathrm{\mathit{\mbox{\ensuremath{p(x)dx},}}}}}}}\label{nxNS}
\end{split}
\end{equation}
where
\[
\Pr\left(\left.x_{2}<\frac{x}{\alpha}\right|x_{1}<\frac{x}{\alpha}\right)=\frac{\int_{0}^{\frac{x}{\alpha}}\int_{0}^{\frac{x}{\alpha}}p(x_{1},x_{2})dx_{1}dx_{2}}{\int_{0}^{\frac{x}{\alpha}}p(x_{1})dx_{1}},
\]
 and
\[
\Pr\left(\left.x_{2}<\frac{x}{\alpha}\right|x_{1}=x\right)=\frac{\int_{0}^{\frac{x}{\alpha}}p(x,x_{2})dx_{2}}{p(x)}.
\]
Then, the return period $R(\alpha,N_{a})$
of unexpected wave crests follows from~Eqs.~\eqref{na}~and~\eqref{NR}. Clearly, if successive waves were stochastically independent, $p(x_{1},x_{2})=p(x_{1})p(x_{2})$ and Eq.~\eqref{nxNS} reduces to~Eq.~\eqref{nx} for the stationary case. 

The theoretical probability structure of two consecutive wave crests is known
for Gaussian processes and it is given by the bivariate Rayleigh distribution~\citep{Fedele2005} 
\begin{equation}
p_{R}(x_{1},x_{2})=256\frac{x_{1}x_{2}}{1-k^{2}}\exp\left[-8\frac{x_{1}^{2}+x_{2}^{2}}{1-k^{2}}\right]I_{0}\left(16 k\frac{x_{1}x_{2}}{1-k^{2}}\right),\label{PW}
\end{equation}
where $I_{0}(y)$ is the modified Bessel function and the parameter $k=\psi\left(\tau_{2}^{*}\right)=\psi_{2}^{*}$
with $\tau_{2}^{*}$ the abscissa of the second absolute maximum of
the normalized covariance function $\psi(\tau)$ of the zero-mean random wave
process~\citep{Fedele2005}. Further, the marginal pdf 
\[
p_{R}(x_{1})=\int_{0}^{\infty}p_{W}(x_{1},x_{2})dx_{2}=16 x_{1}\exp\left(-8 x_{1}^{2}\right)
\]
is the univariate Rayleigh distribution. As $\psi_{2}^{*}$ tends to zero, successive
crests become stochastically independent and the sea state tends to be broadbanded. Thus, we expect that stochastic dependence of waves is dominant
in very narrowband sea states, where $\psi_{2}^{*}\rightarrow1$. In
particular, our numerical simulations discussed later in section~\ref{sec:verify} suggest that the dependence of consecutive crests in
Gaussian seas is dominant when $\psi_{2}^{*}>0.7$. This condition corresponds to unrealistic oceanic
sea states characterized by a Jonswap spectrum with a peak enhancement
factor $\gamma>100$ and very narrowbanded as the spectral bandwidth
$\nu<0.1$. For typical oceanic seas, $\nu\sim0.3-0.5$ and $\psi_{2}^{*}\sim0.2-0.5$, and successive waves can be assumed as stochastically independent.

Note that, the joint pdf of consecutive Gaussian wave crests in Eq.~\eqref{PW} can be generalized to account for second-order bound nonlinearities following~\cite{Fedele2009}, but this is beyond the scope of this work. Since second-order bound harmonics are phase-locked
to the Fourier components of the linear free surface, we expect
the classical Tayfun's~(1980) enhancement of successive linear crest amplitudes, but their dependence
should be unaffected by second-order nonlinearities. 

\subsection{Nonstationarity}

The statistics of unexpected waves formulated so far is valid for
stationary sea states. In nonstationary seas, as those during
storms, our present theory can be formalized as follows. From~Eq.~\eqref{pdfh}, the pdf of an unexpected wave crest height $h$ generalizes to
\begin{equation}
\begin{split}
p_{h}(x;\alpha,N_{a})_{NS}=\idotsint p_{h}(\left.x;\alpha,N_{a}\right|b_{1},\dots b_{M})
\\
\cdot p\left(b_{1},\dots b_{M}\right)\mathrm{d}b_{1}\dots\mathrm{d}b_{M},\label{NS1}
\end{split}
\end{equation}
where $\left\{ b_{j}\right\} _{j=1,M}$ are $M$ time-varying wave
parameters, e.g. $\sigma$,$\mu$,$\lambda_{40}$,$\lambda_{22}$
and $\lambda_{04}$, the conditional pdf $p_{h}(\left.x;\alpha,N_{a}\right|b_{1},\dots b_{M})$
is the stationary pdf in Eq.~\eqref{pdfh} for given values of $b_{j}$
and $p\left(b_{1},\dots b_{M}\right)$ is the joint pdf of the parameters,
which encodes their time variability. Eq.~\eqref{NS1} can be interpreted
as the average value of $p_{h}(\left.x;\alpha,N_{a}\right|b_{1},\dots b_{M})$
with respect to the random variables $b_{j}$, that is
\[
p_{h}(x;\alpha,N_{a})_{NS}=\overline{p_{h}(\left.x;\alpha,N_{a}\right|\mathbf{b})}^{\mathbf{b}},
\]
where the vector $\mathbf{b}=\left[b_{1},\dots b_{M}\right]$ and the
labeled overbar denotes statistical average with respect to $\mathbf{b}$
only. Taylor-expanding around the mean $\overline{\boldsymbol{\mathbf{b}}}=\left[\overline{b}_{1},\dots\overline{b}_{M}\right]$,
up to second order, yields 
\begin{equation}
\begin{split}
p_{h}(x;\alpha,N_{a})_{NS}\simeq\begin{array}{c}
\overline{p_{h}(\left.x;\alpha,N_{a}\right|\overline{\boldsymbol{\mathbf{b}}})}^{\mathbf{b}}+\overline{\sum_{j}\mathbf{g}^{T}\left(\mathbf{b}\mathrm{-}\overline{\boldsymbol{\mathbf{b}}}\right)}^{\mathbf{b}}+
\\
\\
\overline{\left(\mathbf{b}\mathrm{-}\overline{\boldsymbol{\mathbf{b}}}\right)^{T}\mathbf{H}(\overline{\mathbf{b}})\left(\mathbf{b}\mathrm{-}\overline{\boldsymbol{\mathbf{b}}}\right)}^{\mathbf{b}}\end{array},\label{NS2}
\end{split}
\end{equation}
where the superscript $T$ denotes matrix transposition, the vector $\mathbf{g}$ has entries
\[
\mathbf{\left[g\right]_{\mathit{j}}}=\left.\frac{\partial p_{h}(\left.x;\alpha,N_{a}\right|b_{1},\dots b_{M})}{\partial b_{j}}\right|_{\mathbf{b}=\overline{\boldsymbol{\mathbf{b}}}}
\]
and the Hessian matrix
\[
\left[\mathbf{H}(\overline{\mathbf{b}})\right]_{rs}=H_{rs}=\left.\frac{\partial^{2}p_{h}(\left.x;\alpha,N_{a}\right|b_{1},\dots b_{M})}{\partial_{b_{r}}\partial_{b_{s}}}\right|_{\mathbf{b}=\overline{\boldsymbol{\mathbf{b}}}}.
\]
Taking the averages in Eq.~\eqref{NS2} yields
\begin{equation}
\begin{split}
p_{h}(x;\alpha,N_{a})_{NS}\simeq\begin{array}{c}
p_{h}(\left.x;\alpha,N_{a}\right|\overline{b}_{1},...\overline{b}_{M})+
\\
\\
\sum_{rs}\left[\frac{\partial^{2}p_{h}(\left.x;\alpha,N_{a}\right|b_{1},\dots b_{M})}{\partial_{b_{r}b_{s}}}\right]_{\mathbf{b}=\overline{\boldsymbol{\mathbf{b}}}}B_{rs}\end{array},\label{NS3}
\end{split}
\end{equation}
where 
\[
B_{rs}=\overline{\left(b_{r}-\overline{b}_{r}\right)\left(b_{s}-\overline{b}_{s}\right)}^{\beta}
\]
are correlation coefficients, in particular $B_{rr}=\sigma_{b_{r}}^{2}$
is the variance of $b_{r}$. These can be easily estimated from the
nononstationary time series. Thus, $p_{h}$ is the sum of i) the pdf
in Eq.~\eqref{pdfh} evaluated using the mean parameters $\overline{\mathbf{b}})$ and ii) additional terms that account for the spreading of the parameters
from their mean. A similar formula can be obtained for $n(\alpha,N_a)$ in~Eq.~\eqref{nx}. The statistical moments of $p_{h}$ can then be obtained
by integrating Eq.~\eqref{NS3} and the nonstationary return period $N_R^{NS}$ follows from~Eq.~\eqref{NR}.

In our applications (see section~\ref{sec:verify}), time wave measurements
at a point are subdivided in a sequence of optimal $30$-min intervals
during which the sea state can be assumed as stationary. We observed
that shorter time intervals lead to unstable estimates of higher order
moments, whereas longer intervals violate the stationarity assumption.
The variability of the standard deviation $\sigma$ was taken into
account by normalizing the surface height measurements in each 30-min
interval by the respective observed $\sigma$. In our data analysis, wave parameters are estimated as the average values over the available time record. Then, the statistics of unexpected waves can be based on Eq.~\eqref{NS3}, where the $B_{rs}$ terms accounting for non-stationarity are neglected. 

\begin{figure}[th]
\centering\includegraphics[width=1\columnwidth]{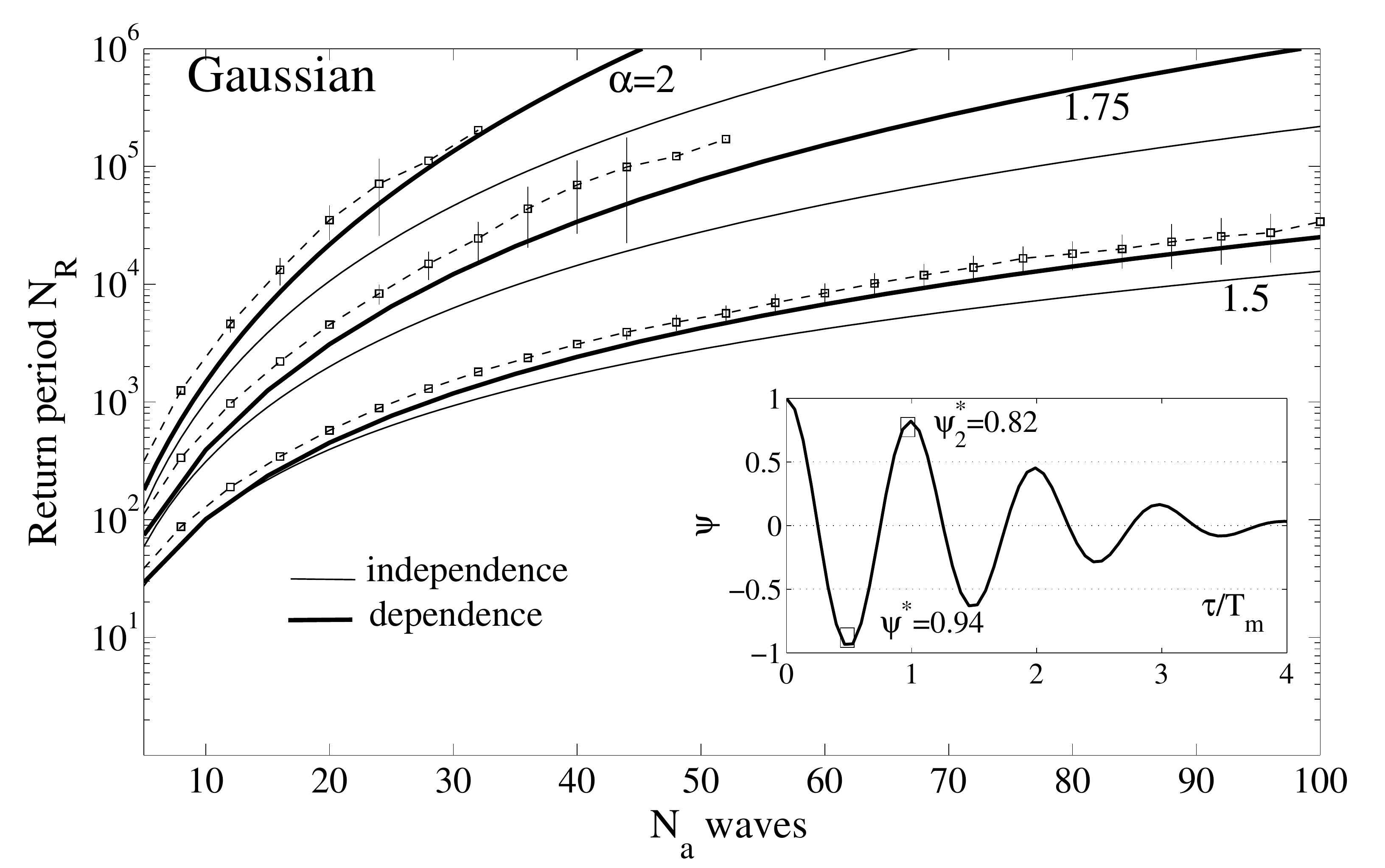}
\protect\caption{The role of stochastic wave dependence to the unexpectedness of crest heights in narrowband Gaussian seas: (thin dashed line with $\square$) empirical one-sided unexpected wave statistics versus (solid lines)
predicted theoretical unconditional return periods $N_{R}$ in number of waves for (thin line) independent and (thick line) dependent crest heights of a wave whose crest height
is $\alpha$-times larger than the surrounding $N_{a}$ waves; $\alpha=1.5,1.75$ and $2$. Confidence bands are also shown. Sea state parameters: Gaussian spectrum with spectral bandwidth $\nu=0.1$ (similar to a Jonswap spectrum with peak enhancement factor $\gamma\sim300$), mean period $T_{m}=8.3$ s, Boccotti parameters $\psi^{*}=0.94$, $\psi^{*}_2=0.82$ and simulated $\sim10^{6}$ waves.}\label{FIG5} 
\end{figure}

\begin{figure}[th]
\centering\includegraphics[width=1\columnwidth]{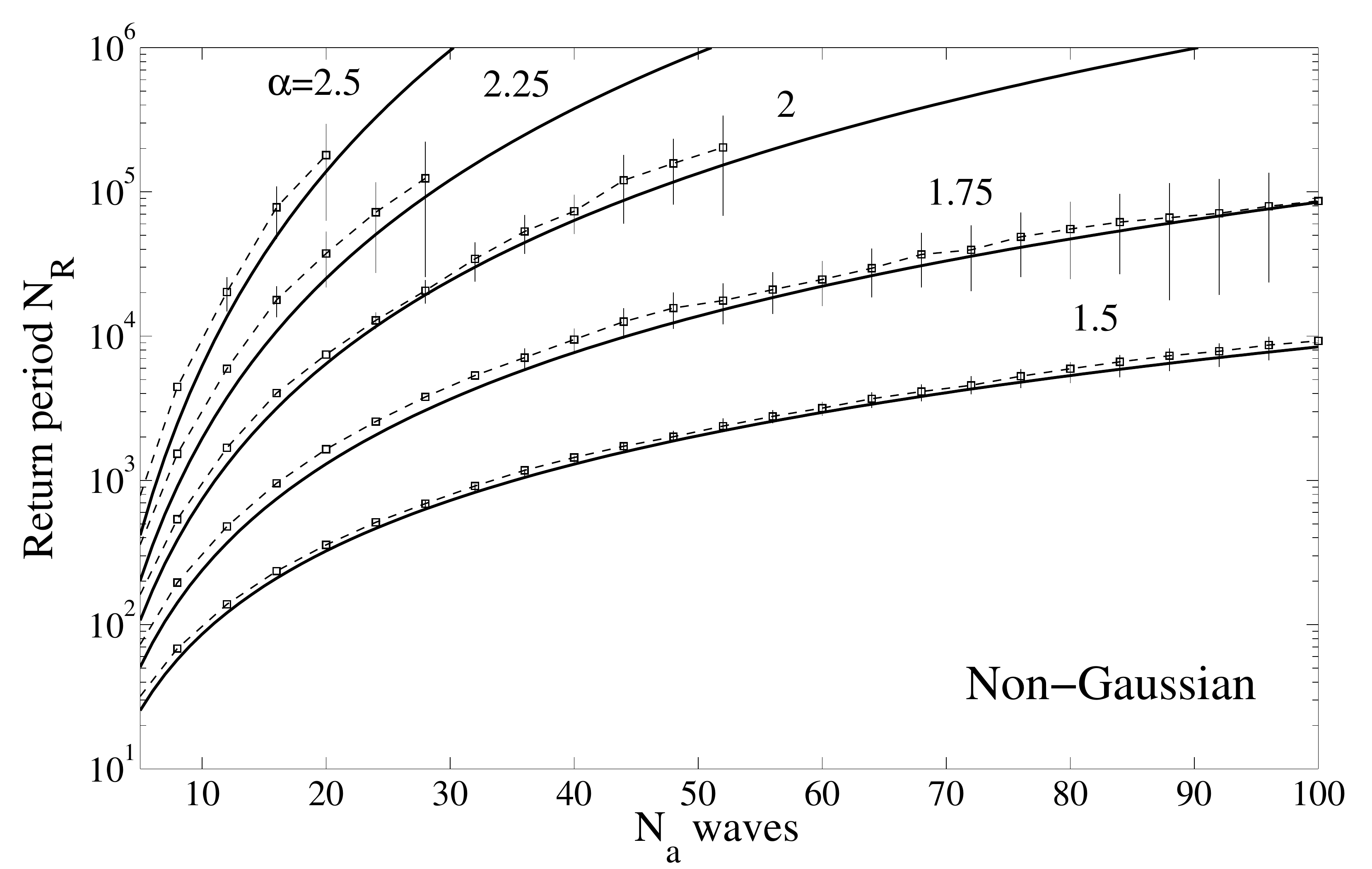}
\protect\caption{Unexpected crest heights in unidirectional second-order random seas. Empirical one-sided (thin dashed lines with~$\square$) unexpected wave statistics versus (thick solid lines) predicted theoretical unconditional return period $N_{R}$ in number of waves of a wave whose crest height
is $\alpha$-times larger than the surrounding $N_{a}$ waves for
increasing values of $\alpha=1.5,1.75,2,2.25$ and $2.5$. Confidence bands are also shown. Sea state parameters: fully developed JONSWAP spectrum (peak enhancement factor $\gamma=1$), mean period $T_{m}=8.3$ s, spectral bandwidth $\nu=0.35$, Tayfun steepness $\mu_m=0.06$ and simulated $\sim10^{6}$ waves. The theoretical predictions accounting for the stochastic independence and dependence of successive crest heights are practically the same as the sea state is broadbanded.}\label{FIG5a} 
\end{figure}

\section{Are rogue waves really unexpected?}

Our interest is to describe statistically the occurrence
of rogue waves with crest heights larger than~$1.25H_{s}$~\citep{DystheKrogstad2008}.
For example, observations indicate that the Andrea rogue wave appeared without
warning suddently, attained a crest height $h_{obs}=1.62H_{s}$, and it was as nearly two-times larger than the surrounding $\mathrm{O}(30)$ waves~\citep{Magnusson2013}. Thus, the Andrea wave is unexpected in accordance with the definition of~\cite{Gemmrich2008}. However, as it will be discussed later in section~\ref{sec:Andrea}, an application of our present theory using Eq.~\eqref{NR} predicts that a wave with a crest height at least twice as that of any of the surrounding $N_{a}=30$ waves occurs on average once every
$N_{R}\sim10^{4}$ waves. This is clearly observed in the left panel of Fig.~\ref{FIG9}. 
Further, the right panel of the same Figure shows that the actual Andrea crest height is nearly the same as the threshold $h_{0.3\cdot 10^6}\sim1.6H_s$ exceeded by the $1/(0.3\cdot 10^{6})$ fraction of largest crests. Eq.~\eqref{Nh} also suggests that the Andrea wave is likely a rare event as the crest threshold $1.6H_{s}$ is exceeded once every $N_h=0.3\cdot 10^{6}$ waves on average. In contrast, our present theory predicts that the Andrea event would occur relatively often as an unexpected wave, i.e. on average once every $N_{R}\sim10^{4}$ waves.

The difference in occurrence rates is explained by first noting that the return period $N_{R}$ is the average time interval between two consecutive waves whose crest height $h$,~\textit{of any possible
amplitude}, is $\alpha$-times larger than the surrounding
$N_{a}$ wave crests. In other words, Eq.~\eqref{harmonicmean} reveals that $N_{R}$ is the harmonic mean of the return periods of all unexpected waves of any crest amplitude, and it is smaller than the return period of large (rare) unexpected waves. Thus, unexpected waves as defined by~\cite{Gemmrich2008} occur relatively often with small or moderate amplitude. However, unexpected waves that are rogue are rare, in agreement with their numerical predictions~(see also~\cite{Gemmrich2010}). 

To quantify the difference in occurrence frequencies of small and large unexpected waves, it is natural to define the conditional return period $N_{R}(\xi,\alpha,N_{a})$
of an unexpected wave whose crest height $h$ exceeds the threshold $\xi H_{s}$ \textit{and} it is $\alpha$-times larger than the surrounding $N_{a}$ wave crests. This is given by 
\begin{equation}
N_{R}(\xi;\alpha,N_{a})=\frac{1}{\int_{\xi}^{\infty}n(x;\alpha,N_{a})dx}=\frac{N_{R}\left(\alpha,N_{a}\right)}{P_{h}(\xi;\alpha,N_{a})},\label{NR-1}
\end{equation}
where 
\begin{equation}
P_{h}(x;\alpha,N_{a})=\int_{x}^{\infty}p_{h}(s;\alpha,N_{a})ds\label{Ph-1}
\end{equation}
is the exceedance probability of the unexpected crest height $h$
from Eq. (\ref{pdfh}). Clearly, for given $\alpha$ and $N_{a}$ the conditional return period $N_{R}(\xi)$ is always greater than the unconditional $N_{R}$ for any $\xi>0$, and they are the same if $\xi=0$. 
The left panel of Fig.~\ref{FIG9} shows that the Andrea rogue wave as an unexpected wave that exceeds $\xi H_s=1.6 H_s$ would occur rarely, i.e. on average once every $N_{R}(\xi=1.6)\sim 6\cdot 10^{6}$. Instead, unexpected waves of any amplitude occur more often, and on average once every $N_{R}\sim10^{4}$. 

Clearly, the Andrea wave is both rogue and unexpected, i.e. its crest is larger than the crests of surrounding waves and it exceeds the threshold~$1.25H_s$~\citep{DystheKrogstad2008}. What is the occurrence frequency of such a bivariate event in comparison to being only rogue as an univariate event?

From Eq.~\eqref{NR-1} the following inequality holds
\begin{equation}
N_{R}(\xi)\geq \frac{1}{\int_{\xi}^{\infty} p(x) dx}=\frac{1}{P(\xi)}=N_h(\xi),
\end{equation}
where we have used $n(x;\alpha,N_{a})\leq p(x)$ from Eq.~\eqref{nx}. Here, $N_h(\xi)$ is defined in~Eq.~\eqref{Nh}) as the standard unconditional return period (in number of waves) of a wave whose crest exceeds the threshold $h=\xi H_s$. 
Thus, a wave whose crest is both larger than $\xi H_s$ and unexpected (as being larger than the surrounding waves) has a lower occurrence frequency than a wave whose crest is just larger than the same threshold. 

The preceding results imply that a rogue wave that is also unexpected has a lower occurrence frequency than just being rogue. For example, for the Andrea sea state the return period of a crest larger than $h_n=1.6H_s$  is $N_h(h_n)=0.3\cdot 10^{6}$. This is smaller than the return period $N_{R}$ of an unexpected wave exceeding the same threshold, i.e. $N_{R}(\xi=1.6)\sim 6\cdot 10^{6}$~(see left panel of Fig.~\ref{FIG9}). Similar conclusions hold for the WACSIS rogue wave (see section~\ref{sec:verify}). 

\begin{figure*}[th]
\centering\includegraphics[width=1\textwidth]{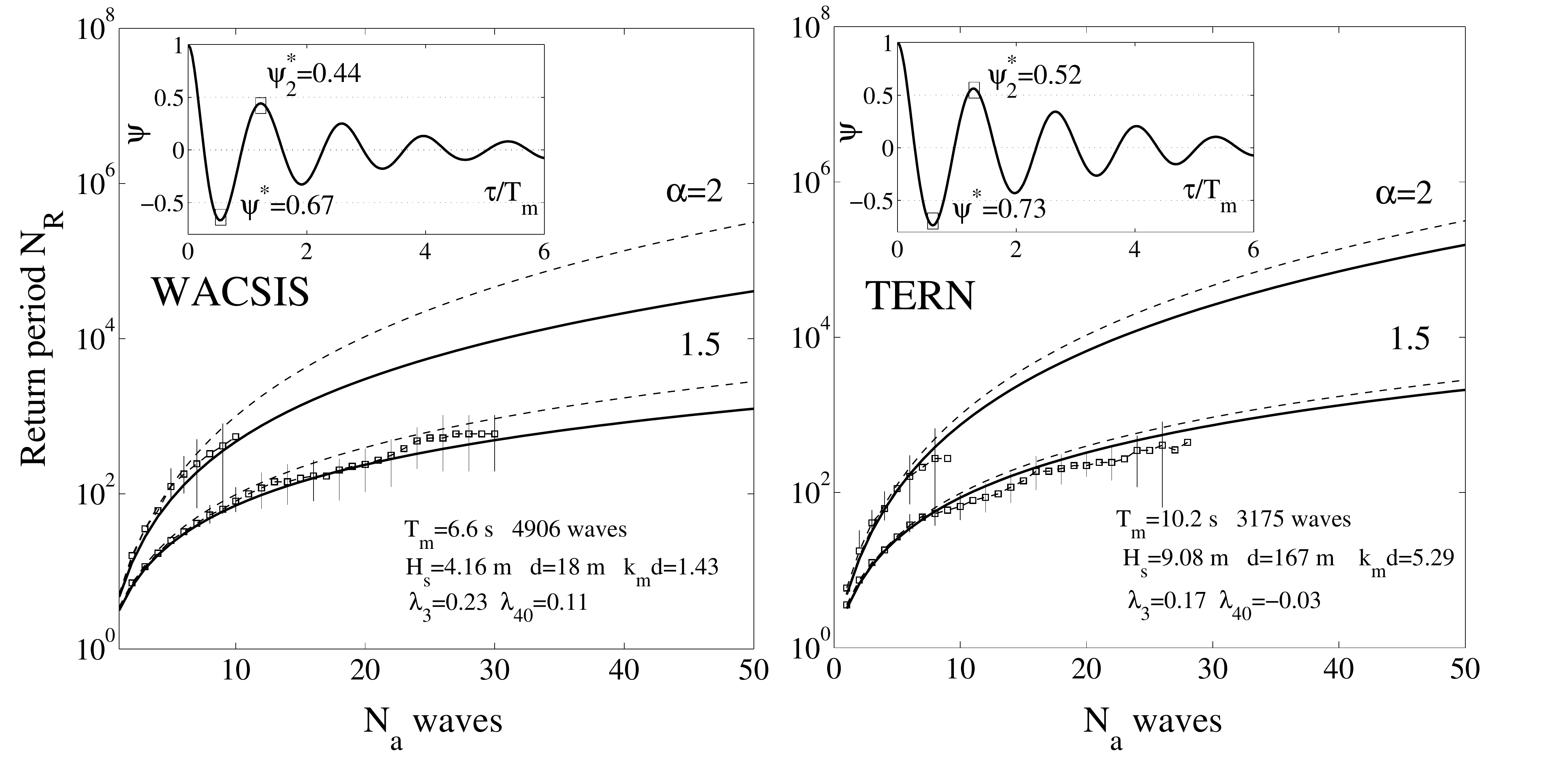}
\protect\caption{Left panel: WACSIS, predicted theoretical nonlinear unconditional return period $N_{R}$ in number of waves (solid line) of a wave whose crest height is $\alpha$-times larger than the surrounding $N_{a}$ waves, linear predictions (dash lines) and
empirical one-sided observed statistics ($\square$) for $\alpha=1.5$ and $2$. Confidence bands are also shown. Right panel: same for TERN measurements. Statistical parameters are taken from~\cite{Tayfun2006,TayfunFedele2007}.} \label{FIG6} 
\end{figure*}

\section{Verification and comparisons}\label{sec:verify}

\subsection{Monte Carlo simulations of Gaussian seas}

Drawing on~\cite{Gemmrich2008}, we performed Monte Carlo simulations of a Gaussian sea described by the average JONSWAP spectrum with peak enhancement factor $\gamma=1$. The sea state is broadbanded with mean period $T_m=8.3$ s, peak period $T_p=10$ s, spectral bandwidth $\nu\sim0.35$ and Boccotti parameters $\psi^{*}=0.65$, $\psi^{*}_2=0.3$ (see covariance function in the panel inset of Fig.~\ref{FIG2}).  A long time series of wave surface displacements was randomly generated containing a total of $\sim10^{6}$ waves, from which unexpected waves were sampled. As the sea state is broadbanded, our theoretical predictions can be based on Eqs.~\eqref{na}~and~\eqref{NR} assuming the stochastic independence of successive crest heights. 

The left panel of Fig.~\ref{FIG2} shows the empirical return period $N_{R}=R/T_{m}$
in number of waves of both one-sided (thin dashed line) and two-sided
(thin solid line) unexpected wave crests as a function of the surrounding
$N_{a}$ waves for different values of $\alpha$ ($N_{a}$ even for the two-sided statistics). 
The two statistics are roughly the same with two-sided unexpected waves slightly less frequent than the one-sided waves. Note that for the two-sided unexpectedness~\cite{Gemmrich2008} consider $N_a$ waves on each side, thus their two-sided return period is larger than ours. 
Shown in the right panel of Fig.~\ref{FIG2} are also the empirical statistics of mean crest heights in comparison to our theoretical predictions for stochastically independent waves. In particular, we note that the mean crest height of two-sided unexpected waves is slightly smaller than that of one-sided waves, especially as $\alpha$ increases. Further, in Fig.~\ref{FIG3} there are shown the predicted conditional return periods $N_R(\xi)$ (solid lines) of an unexpected wave whose crest height is greater than $\xi H_s$  for $\xi=0,1.0$ and $1.2$~$(\alpha=1.5)$. Note that $N_R(\xi=0)$ is the unconditional return period $N_R$. We find a fair agreement with the empirical one-sided unexpected wave statistics (squares). 
For $\alpha=2$ and $N_a=30$ our predicted return period is $N_R\sim6\cdot10^4$ and in fair agreement with the linear predictions ($\sim 7\cdot10^4$) by~\cite{Gemmrich2008} as shown in their Fig.~2. As regard to unexpected crest-to-trough heights, our theoretical model fairly predicts the empirical wave height statistics from simulations as clearly seen in Fig.~\ref{FIG4}. 

In the above comparisons, the fair agreement with our theoretical predictions indicates that the stochastic independence of waves holds approximately as the sea state is broadbanded. However, in very narrowband seas the stochastic dependence of neighboring waves cannot be neglected. 
Indeed, consider a linear sea state characterized with a Gaussian spectrum with spectral bandwidth $\nu=0.1$. This is similar to an unrealistic Jonswap spectrum with peak enhancement factor $\gamma\sim300$. From the panel inset of Fig.~\ref{FIG5}, the Boccotti parameters are $\psi^{*}=0.94$ and $\psi^{*}_2=0.81$ indicating a strong correlation between consecutive waves. Indeed, from the same figure the empirical one-sided (square) unexpected wave statistics tends to agree with our predicted theoretical return period $N_{R}$ for dependent waves (thick solid line) computed using Eqs.~\eqref{nxNS}~and~\eqref{NR}. Instead, our predictions for independent waves (thin solid line) are less conservative, where we use Eqs.~\eqref{na}~and~\eqref{NR}.

\begin{figure}[th]
\centering\includegraphics[width=0.8\columnwidth]{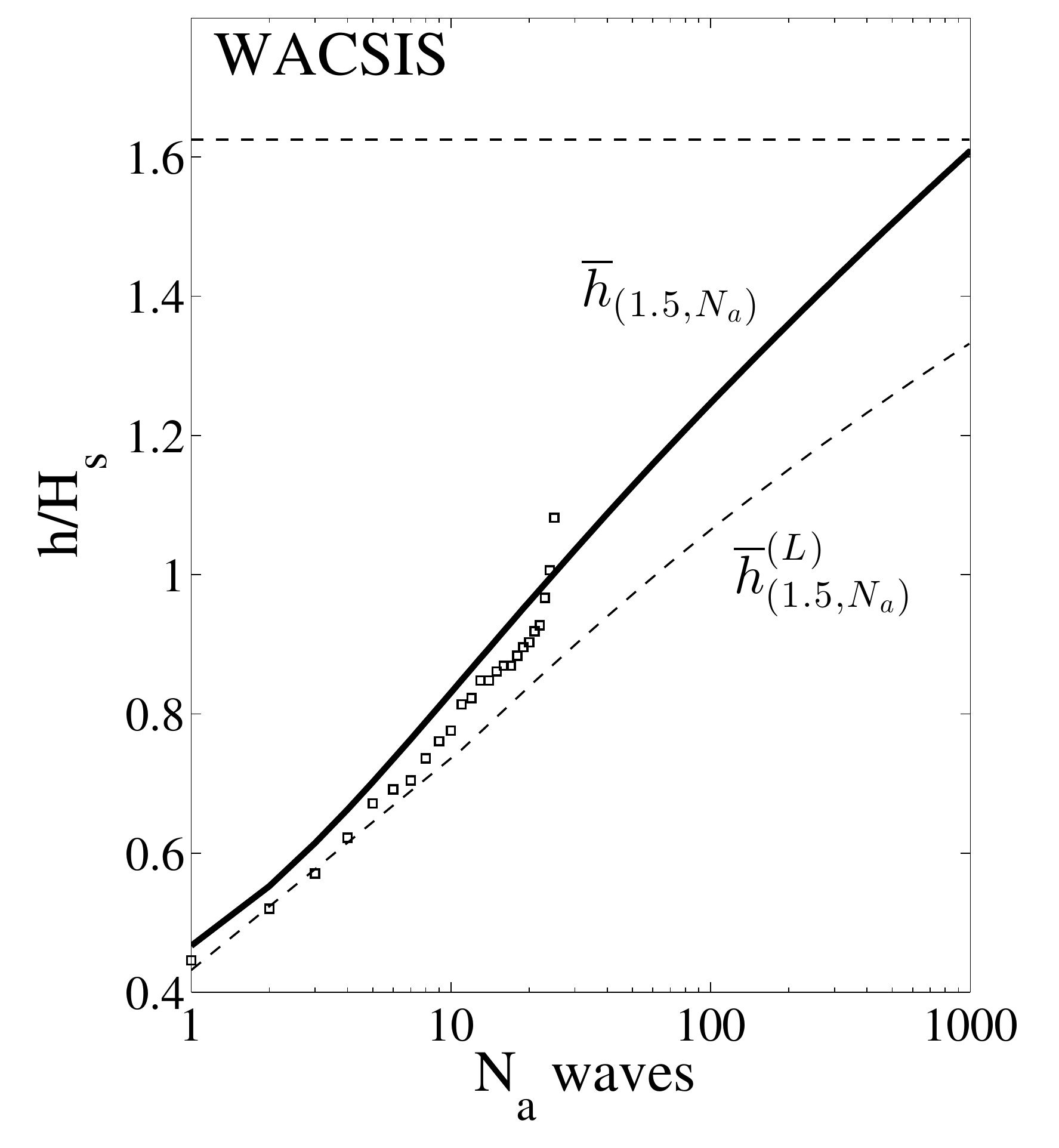}
\protect\caption{WACSIS unexpected wave crest heights: predicted theoretical nonlinear (solid line)
and linear (dash line) mean heights $\overline{h}_{\alpha,N_{a}}$
and $\overline{h}_{\alpha,N_{a}}^{(L)}$ as a function of the number
$N_{a}$ of surrounding waves versus empirical one-sided statistics
(squares) for $\alpha=1.5$. Horizontal line denotes the observed maximum crest height $1.62H_{s}$. Wave parameters $H_{s}=4.16$ m, $T_{m}=6.6$ s, depth $d=18$ m~\citep{WACSIS2002}. Average wave parameters are taken from~\cite{Tayfun2006,TayfunFedele2007}, in particular skewness $\lambda_3\sim 0.23$ and excess kurtosis $\lambda_{40}\sim 0.11$.}
\label{FIG7} 
\end{figure}

\subsection{Monte Carlo simulations of second-order random seas}

Drawing on~\cite{TayfunFedele2007}, we performed Monte Carlo simulations of unidirectional second-order broadband random seas in deep water described by the same average JONSWAP spectrum introduced in the previous section for simulating Gaussian seas. The associated Tayfun~(1980)~steepness $\mu=\lambda_3/3\sim0.06$, where $\lambda_3$ is the skewness of surface elevations~\citep{Fedele2009}. 
Our theoretical predictions are based on Eqs.~\eqref{na}~and~\eqref{NR} and assume the stochastic independence of successive crest heights as the sea state is broadbanded.

In Fig.~\ref{FIG5a} it is shown the comparison between the empirical return period $N_{R}=R/T_{m}$
in number of waves of one-sided (squares) unexpected wave crests and theoretical predictions from our model as a function of the surrounding $N_{a}$ waves for different values of $\alpha$. 
For $\alpha=2$ and $N_a=30$  our predicted second-order return period $N_R\sim2\cdot10^4$ is shorter than the linear counterpart ($\sim6\cdot10^4$) for Gaussian seas~(see  Fig.~\ref{FIG2}) as nonlinearities enhance crest heights~\citep{TayfunFedele2007,Fedele2009}. Further, our second-order predictions fairly agree with those by~\cite{Gemmrich2008} in their Fig.~2. For example, they predict a slightly shorter nonlinear period $N_R\sim10^4$ for $\alpha=2$ and $N_a=30$. This is because their second-order correction for crest heights is based on the narrowband assumption of the sea state. This yields a slightly overestimation of crest heights shortening $N_R$. In contrast, our simulated sea states are based on the exact second-order solution for unidirectional broadband waves in deep water~\citep{Tayfun1980}. 

\subsection{Oceanic observations}

We will analyze two data sets. The first comprises 9 h of measurements
gathered during a severe storm in January, 1993 with a Marex radar
from the Tern platform located in the northern North Sea in a water
depth of $d=167$ m. We refer to \cite{Forristall2000} for further
details on the data set, hereafter referred to as TERN. The second
data set is from the Wave Crest Sensor Intercomparison Study (WACSIS)
(\cite{WACSIS2002}). It consists of 5 h of measurements gathered
in January, 1998 with a Baylor wave staff from Meetpost Noordwijk
in the southern North Sea (average water depth $d=18$ m). \cite{Tayfun2006} and \cite{TayfunFedele2007} elaborated both data sets and provided accurate estimates of statistical parameters, especially skewness and fourth-order cumulants which will be used in this work.
The data analysis indicates that the statistics of unexpected waves can be based on Eq.~\eqref{NS3}, where the $B_{rs}$ terms accounting for non-stationarity are neglected. 
Further, successive waves can be assumed as stochastically independent as the both sea states are broadbanded as indicated by their estimated covariance functions~(see panel insets in~Fig.~\ref{FIG6}). 

\begin{figure*}[th]
\centering\includegraphics[width=1\textwidth]{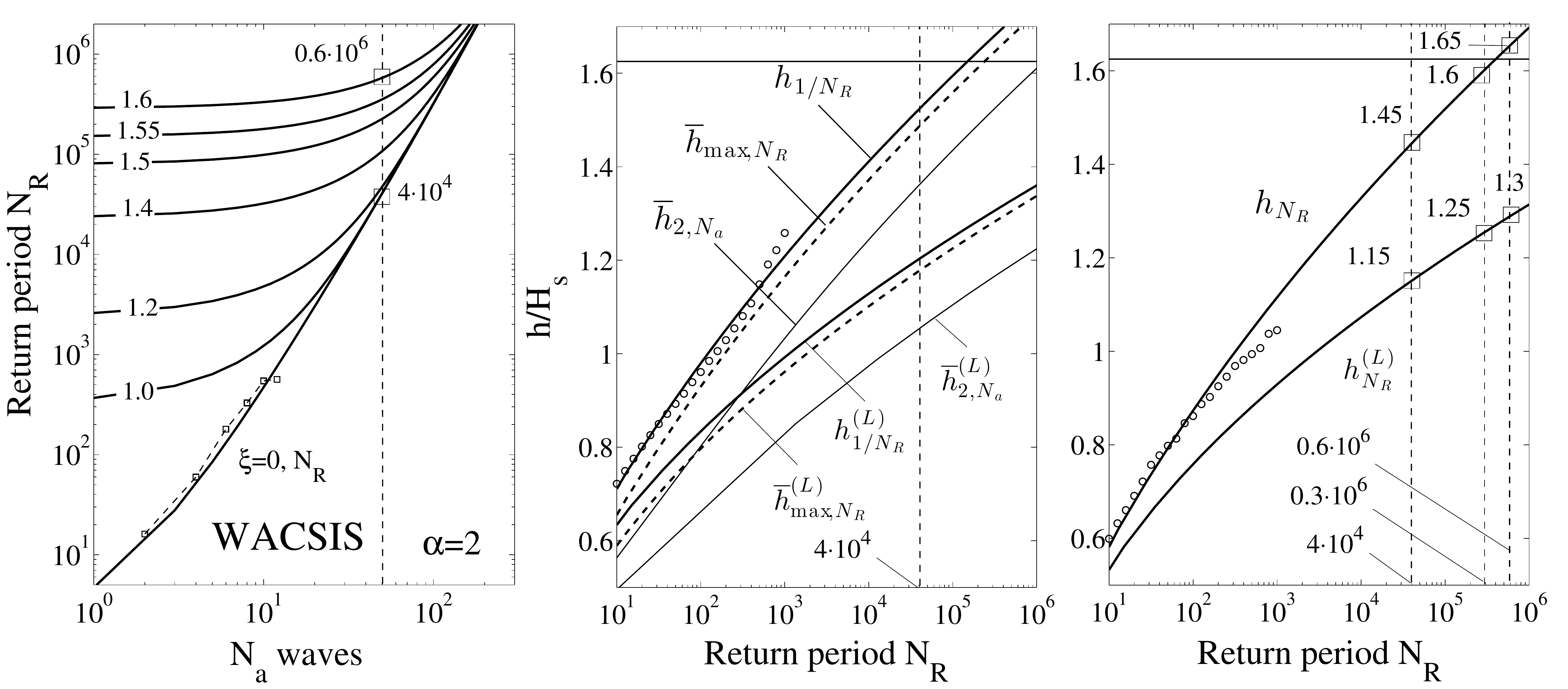}
\protect\caption{WACSIS rogue wave: (Left panel) predicted nonlinear theoretical return periods $N_R(\xi)$, in number of waves, of unexpected crest heights greater than $\xi H_s$ and $\alpha=2$-times larger than the surrounding $N_{a}$ waves for $\xi=0,1.0,1.2,1.4,1.55$ and $1.6$ (solid lines) and (square) empirical one-sided unexpected wave statistics. Dashed vertical line denotes return period values at $N_a=50$. (Center panel) predicted nonlinear mean crest height $\overline{h}_{\max,N_{R}}$, conditional mean $h_{1/N_{R}}$ and average unexpected crest height $\overline{h}_{\alpha=2,N_a}$ versur their linear counterparts as a function of number of waves $N_{R}$. Empirical conditional mean $h_{1/N_{R}}$ is also shown (circles). (Right panel) predicted (solid line) and empirical (circles) nonlinear threshold $h_{N_{R}}$ versus its linear counterpart as a function of $N_{R}$. Dashed vertical lines denote values at $N_R=4\cdot 10^4,0.3\cdot10^6$ and $0.6\cdot 10^6$. The horizontal line denotes the observed maximum crest height $1.62H_{s}$. Average wave parameters are taken from~\cite{Tayfun2006,TayfunFedele2007}, in particular skewness $\lambda_3\sim 0.23$ and excess kurtosis $\lambda_{40}\sim 0.11$.}\label{FIG8} 
\end{figure*}

\begin{figure*}[th]
\centering\includegraphics[width=1\textwidth]{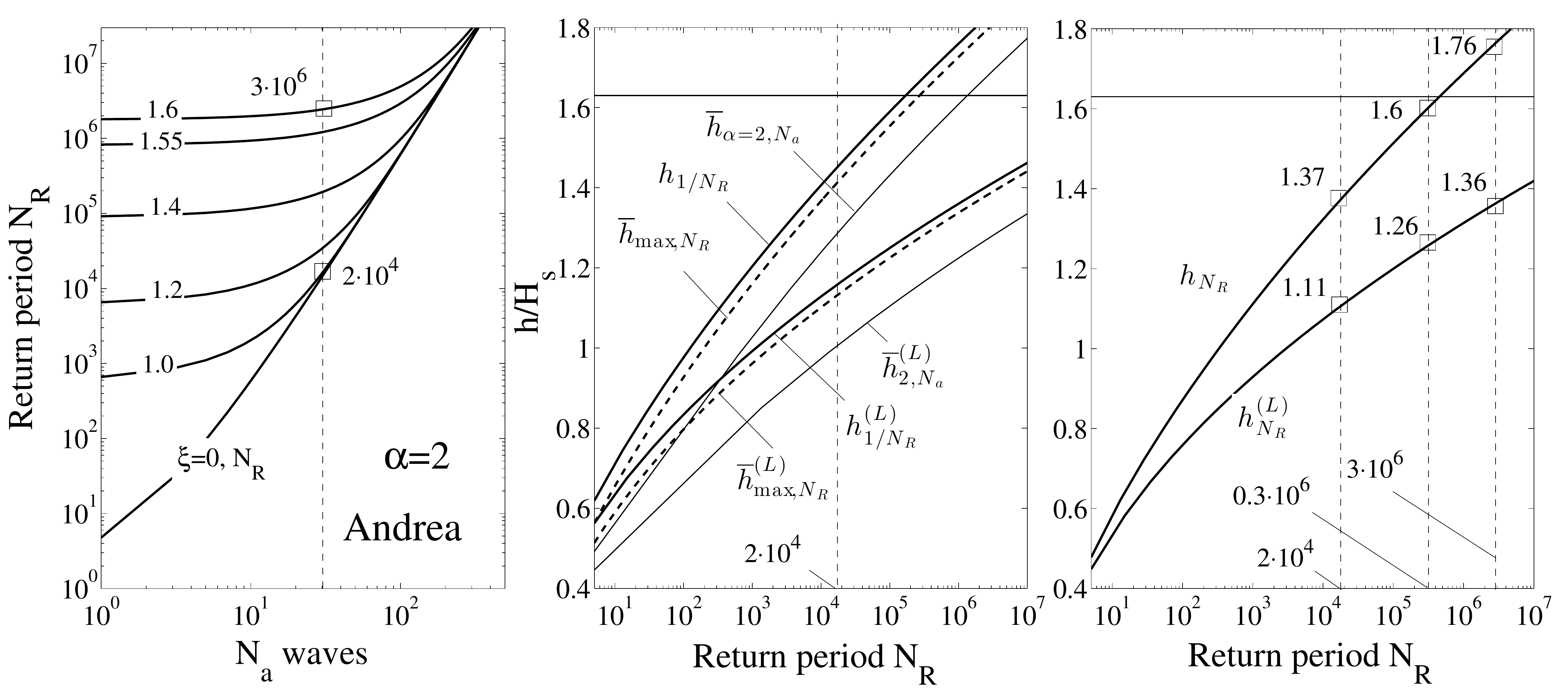} \protect\caption{Andrea rogue wave: (Left panel) predicted nonlinear theoretical return periods $N_R(\xi)$, in number of waves, of unexpected crest heights greater than $\xi H_s$ and $\alpha=2$-times larger than the surrounding $N_{a}$ waves for $\xi=0,1.0,1.2,1.4,1.55$ and $1.6$. Dashed vertical line denotes return period values at $N_a=30$. (Center panel) predicted nonlinear mean crest height $\overline{h}_{\max,N_{R}}$, conditional mean $h_{1/N_{R}}$ and average unexpected crest height $\overline{h}_{\alpha=2,N_a}$ versus their linear counterparts as a function of number of waves $N_{R}$. Empirical conditional mean $h_{1/N_{R}}$ is also shown (circles). (Right panel) predicted (solid line) and empirical (circles) nonlinear threshold $h_{N_{R}}$ versus its linear counterpart as a function of $N_{R}$. Dashed vertical lines denote values at $N_R=2\cdot 10^4,0.3\cdot 10^6$ and $3\cdot 10^6$. The horizontal line denotes the observed maximum crest height $1.63H_{s}$. Wave parameters $H_{s}=9.2$ m, $T_{m}=13.2$ s, depth $d=70$ m \citep{Magnusson2013}, skewness $\lambda_{3}=0.15$ and excess kurtosis $\lambda_{40}=0.1$~\citep{Dias2015}.}\label{FIG9} 
\end{figure*}

As regard to WACSIS measurements, the left panel of Fig.~\ref{FIG6} compares the theoretical nonlinear return period $N_{R}$ (solid line) of unexpected wave crests
$\alpha$-times larger than the surrounding $N_{a}$ waves, the respective linear
predictions (dashed line) and the WACSIS empirical one-sided statistics  for~$\alpha=1.5,2$~(dashed line with $\square$). The right panel of the same figure shows similar comparisons
for TERN. The observed occurrence rates are close to the theoretical predictions, indicating that the assumption of stochastic independence of waves holds approximately. It is noticed that nonlinearities tend to reduce the return period of unexpected waves and increase their mean crest amplitudes. In particular, in the left panel of Fig.~\ref{FIG7} we compare our predicted
nonlinear (solid line) and linear (dash line) mean crest heights $\overline{h}(\alpha,N_{a})$
and $\overline{h}^{(L)}(\alpha,N_{a})$ versus the WACSIS empirical one-sided
statistics ($\square$) for $\alpha=1.5$. 
Clearly, our linear predictions underestimate the observed crest amplitudes, as expected. Indeed, it is well established that nonlinearities must be accounted for to obtain reliable statistics of unexpected waves~\citep{Tayfun1980,Forristall2000,TayfunFedele2007,Fedele2009,GG2011}. Similar trend is also observed for the WACSIS rogue wave as evident from the center panel of Fig.~\ref{FIG8}. Here, there are shown our nonlinear predicted mean crest height $\overline{h}_{\max,N_{R}}$, conditional mean $h_{1/N_{R}}$ and mean unexpected crest height $\overline{h}_{\alpha=2,N_a}$ versus their linear counterparts. The right panel of the same figure depicts the nonlinear threshold $h_{N_R}$ in comparison to its linear counterpart. The nonlinear and linear predictions for the Andrea rogue wave are also shown in~Fig.~\ref{FIG9}.

We observe that the empirical statistics tend to deviate
from the theoretical predictions for large values of $\alpha$ and
$N_{a}$. In particular, for both TERN and WACSIS we
could not produce statistically stable estimates of extreme values for $N_{a}>10$ when $\alpha>1.5$ 
due to the limited number of waves in the time series ($O(10^{3})$ waves in comparison to the $10^{6}$ waves of the simulated Gaussian seas). Nevertheless, the
agreement between our present theory and observations is satisfactory and it also provides evidence that successive waves in the samples are approximately stochastically independent.

\section{How rogue are unexpected waves? }

WACSIS observations indicate that the actual largest crest $h_{obs}$ is $1.62H_s$. Fig.~\ref{FIG1} shows that the WACSIS rogue wave is also unexpected as it is $\alpha=2$-times larger than the surrounding $N_{a}\sim50$ waves. According to our statistical model such unexpected wave would occur often and on average once every $N_{R}=4\cdot 10^{4}$ waves, as seen in the left panel of Fig.~\ref{FIG8}. Here, we report the theoretical predictions of the unconditional nonlinear return period $N_R$ as a function of $N_a$ using Eqs.~\eqref{na}~and~\eqref{NR}). 
Further, from the center panel of Fig.~\ref{FIG8} it is seen that the associated average nonlinear unexpected crest height $\overline{h}_{(\alpha=2,N_a=50)}$ is about $1.35 H_s$ and smaller than the conditional mean $h_{1/N_R}\sim1.5H_s$, which is slightly larger than the mean maximum crest height $\overline{h}_{\max,N_R}=1.48 H_s$ of $N_R=4\cdot 10^{4}$ waves. Note that these average values underestimate the actual maximum crest amplitude $h_{obs}\sim1.62H_{s}$ observed. In contrast, the right panel of Fig.~\ref{FIG8} shows that $h_{obs}$ is nearly the same as the threshold $h_{0.3\cdot^{6}}=1.6H_s$ exceeded on average once every $N_h=0.3\cdot 10^6$ waves~(see~Eq.~\eqref{Nh}). 

We have seen that a correct statistical interpretation of the WACSIS rogue wave as an unexpected event requires considering the conditional return period $N_R(\xi)$ of an unexpected wave whose crest height is larger than $\xi H_s$~(see Eq.~\eqref{NR-1}). In particular, the left panel of~Fig.~\ref{FIG8} depicts plots of $N_R(\xi)$ as a function of $N_a$ for increasing values of $\xi=1,1.2,1.4,1.55$ and $1.6$~$(\alpha=2)$. For $\xi=1.6H_s$, we find that an unexpected wave exceeding this threshold and standing above $N_a=50$ waves would occur rarely and once every $N_{R}(\xi=1.6)\sim0.6\cdot 10^{6}$, in contrast to the smaller unconditional value $N_{R}\sim 4\cdot 10^{4}$. 

In summary, the WACSIS wave crest as both unexpected and rogue, i.e. two-times larger than $N_a=50$ surrounding waves and exceeding the $1.6H_s$, would occur once every $N_R=0.6\cdot10^6$ waves on average. In contrast, the WACSIS wave as a rogue event has a crest height that is nearly the same as the threshold $h_{0.3\cdot10^6}=1.6H_s$ exceeded on average once every $N_h=0.3\cdot10^6$ waves. Thus, the WACSIS rogue wave has a slightly greater occurrence frequency than being both rogue and unexpected since $N_h<N_R=0.6\cdot10^6$. This implies that the threshold $h_{N_R}$ exceeded by the $1/N_R$ fraction of the largest crests is larger than $1.6H_s$ and nearly the same as $1.65H_s$.

\section{The Andrea rogue wave and its unexpectedeness}\label{sec:Andrea}

As a specific application of the present theoretical framework, the unexpected wave statistics of the 2007 Andrea rogue wave event is examined. The actual largest crest height $h_{obs}$ is $1.63H_s$  and nearly two-times larger than the surrounding $\mathrm{O}(30)$ waves (see Fig.~12 in \cite{Magnusson2013}). For the hindcast Andrea sea state, the left panel of Fig.~\ref{FIG9} shows the unconditional and conditional nonlinear return periods $N_R$ and $N_R(\xi)$ as a function of $N_a$. In particular, according to our statistical model, the theoretical predictions indicate that a wave with a crest height at least twice that of any of the surrounding $N_{a}=30$ waves occurs on average once every $N_{R}=2\cdot 10^{4}$ waves irrespective of its crest amplitude. In contrast, an unexpected wave whose crest height exceeds the threshold $1.6H_s$ occurs less often since our predicted conditional return period $N_{R}(\xi=1.6)\sim 3\cdot 10^{6}$ is greater than the unconditional counterpart $N_{R}=2\cdot 10^{4}$, as seen in the left panel of~Fig.~\ref{FIG9}. Furthermore, the crest height $1.6H_s$ is nearly the same as the threshold $h_{0.3\cdot 10^6}$ exceeded on average once every $N_h=0.3\cdot 1/10^6$ waves, as indicated in the right panel of the same figure. Thus, the Andrea wave has a greater occurrence rate than being both rogue and unexpected since $N_h<N_R=3\cdot10^6$, and implying the larger threshold $h_{N_R}=1.76H_s$.

\section{Concluding remarks}

We have presented a third-order nonlinear model for the statistics of unexpected waves. \cite{Gemmrich2008} define as unexpected a wave that is taller than a set of neighboring waves. The term "unexpected" refers to a wave that is not foreseen by a casual observer~\citep{Gemmrich2010}. Clearly, unexpected waves are predictable in a statistical sense. Indeed, they can occur relatively often with a small or moderate crest height. However, unexpected waves that are rogue are rare. This difference in occurrence frequencies is quantified by introducing the conditional return period of an unexpected wave that exceeds a given threshold crest height. The associated unconditional return period is smaller than the conditional counterpart as it refers to the harmonic mean of the return periods of unexpected waves of any crest amplitude. 

Furthermore, our analysis indicate that a wave that is both rogue and unexpected has a lower occurrence frequency than just being rogue. This is proven both analytically and verified by way of an analysis of the Andrea and WACSIS rogue wave events. Both waves appeared without warning and their crests were nearly $2$-times larger than the surrounding $O(10)$ wave crests, and thus unexpected. The two crest heights are nearly the same as the threshold $h_{0.3\cdot10^{6}}\sim1.6H_{s}$ exceeded on average once every $0.3\cdot 10^{6}$ waves. In contrast, the Andrea and WACSIS events would occur less often as being both unexpected and rogue, i.e. on average once every $3\cdot10^{6}$ and $0.6\cdot10^6$ respectively. 

Finally, we point out that our statistical model for unexpected waves supports and goes beyond the analysis by~\cite{Gemmrich2008} based on Monte Carlo simulations. In particular, our statistical approach can be used in operational wave forecast models to predict the unexpectedness of ocean waves.

\section{Acknowledgments}

FF is grateful to George Z. Forristall and M. Aziz Tayfun for sharing the wave measurements utilized in this study. FF thanks Michael Banner, George Forristall, Peter A. E. M. Janssen, Victor Shrira and M. Aziz Tayfun for discussions on nonlinear wave statistics. FF also thanks M. Aziz Tayfun for sharing his numerical solver for simulating second-order nonlinear waves. Further, FF thanks Michael Banner and M. Aziz Tayfun for revising an early draft of the manuscript as well as Guillermo Gallego for his support with \LaTeX{}. FF acknowledges partial support from NSF grant CCF-1347191.

%
%
%
\bibliographystyle{ametsoc2014}
\bibliography{unexpectedwaves}

%

%

\end{document}